\documentclass{article}
\usepackage{spconf,amsmath,graphicx,hyperref,booktabs}

\title{Multi-Stage Music Source Restoration with \\ BandSplit-RoFormer Separation and HiFi++ GAN}
\name{Tobias Morocutti$^\star$, Emmanouil Karystinaios$^\star$\thanks{$\star$: Equal contribution among authors}, Jonathan Greif, Gerhard Widmer}
\address{Institute of Computational Perception (CP-JKU) \\ Johannes Kepler University Linz, Austria \\ 
firstname.lastname@jku.at}
\begin{document}
%
\maketitle
\begin{abstract}
Music Source Restoration (MSR) targets recovery of original, unprocessed instrument stems from fully mixed and mastered audio, where production effects and distribution artifacts violate common linear-mixture assumptions. This technical report presents the CP-JKU team’s system for the MSR ICASSP Challenge 2025. Our approach decomposes MSR into separation and restoration. First, a single BandSplit-RoFormer separator predicts eight stems plus an auxiliary other stem, and is trained with a three-stage curriculum that progresses from 4-stem warm-start fine-tuning (with LoRA) to 8-stem extension via head expansion. Second, we apply a HiFi++ GAN waveform restorer trained as a generalist and then specialized into eight instrument-specific experts.
\end{abstract}

\section{Introduction}

Professional music production violates the linear-mixture assumptions typically used by conventional music source separation methods. Release pipelines commonly include equalization, dynamic range compression, reverberation, saturation and distortion, stereo widening, limiting, and codec artifacts, often compounded by additional degradations. As a result, the target sources are not only mixed but also systematically transformed, making direct separation under clean-stem assumptions insufficient. The MSR Challenge therefore targets recovering the original, unprocessed sources for eight instrument classes (vocals, guitar, keyboard, synthesizer, bass, drums, percussion, orchestra) from such mixtures.

We address MSR with a modular learning setup that explicitly separates the de-mixing problem from the de-mastering and de-artifacting problem. Concretely, we adopt a separation-then-restoration strategy: a single multi-stem separator estimates degraded stems, and per-stem restoration networks map these estimates toward clean targets. This design enables us to train the separator on large-scale augmented mixtures while training restoration models to invert production effects and suppress artifacts under realistic separator error distributions. In this submission, we make two practical contributions: (i) a curriculum for BS-RoFormer adaptation from 4 to 8 stems using parameter-efficient LoRA fine-tuning and head expansion, (ii) instrument-specific restoration experts trained on separator-generated inputs to improve train-test alignment.

\section{System overview}
Our system processes a mixture waveform $x$ via a two-stage pipeline. In the first stage, a single multi-stem separator $\mathcal{S}$ estimates eight target stems and an auxiliary \textit{other} stem:
\begin{equation}
\hat{s}_k = \mathcal{S}(x)_k,\quad k \in \{1,\dots,8\}, \qquad \hat{s}_{\mathrm{other}} = \mathcal{S}(x)_{\mathrm{other}} .
\end{equation}
In the second stage, an instrument-specific restoration expert $\mathcal{R}_k$ maps each separated estimate to a restored stem:
\begin{equation}
\tilde{s}_k = \mathcal{R}_k(\hat{s}_k),\quad k \in \{1,\dots,8\}.
\end{equation}
To improve train-test alignment for restoration, we generate restoration inputs by running the trained separator on our synthetic training mixtures; each expert is then trained to map the resulting separator output to the corresponding clean stem, exposing the restorers to realistic separation errors.\footnote{Code and model details are available at: \\ \url{https://github.com/CPJKU/music-source-restoration}.}

\subsection{Source Separation}


We use a BandSplit-RoFormer (BS-RoFormer) separator~\cite{lu2024music}. The model applies a band-split front-end to process different frequency regions separately and uses RoFormer blocks to model temporal and cross-band dependencies for mask estimation. A single model predicts all stems using 9 mask estimation heads (8 instrument classes plus auxiliary \textit{other}).


We train the separator in three stages, starting from a publicly available 4-stem BS-RoFormer checkpoint:
\begin{itemize}
    \item \textbf{Stage 1} (4 stems, clean mixtures). Fine-tune on clean mixtures (no additional augmentation/mastering) to separate {vocals, drums, bass, other}.
    \item \textbf{Stage 2} (4 stems, mastered mixtures). Continue fine-tuning on mixtures created by an online per-stem degradation pipeline followed by mixing/mastering. The target stems are the degraded stems.
    \item \textbf{Stage 3} (8 stems). Expand the model to 8 stems. Shared layers and the original four heads are initialized from Stage 2. New heads are randomly initialized and only the new mask heads are trained, while the backbone remains frozen.    
\end{itemize}

In Stages 1 and 2, we fine-tune the transformer layers using LoRA. We use a weighted combination of: i) masked SI-SNR loss, ii) multi-resolution STFT loss, iii) L1 loss, and iv) low-amplitude penalty loss.


For the 4-stem stages, we combine MUSDB18-HQ~\cite{musdb18-hq}, DSD100~\cite{SiSEC16}, and MoisesDB; for the 8-stem stage, we combine MoisesDB~\cite{pereira2023moisesdb}, Slakh2100~\cite{manilow2019cutting}, MedleyDB v2~\cite{bittner2016medleydb}, RawStems~\cite{zang2025music}, and MUSDB25-style extensions~\cite{musdb25}. We apply an online augmentation and degradation strategy to each source independently and then apply a mixing and mastering strategy to combine the degraded sources into the final mixture; the separator is trained to predict these degraded stems. 
The model was trained on 4 NVIDIA H100 GPUs.

\subsection{Restoration}

The restoration module is a HiFi++ GAN bundle comprising a SpectralUNet front-end, an upsampling stage, a WaveUNet refinement network, and a SpectralMaskNet for residual spectral correction. We first train a generalist restoration model and then fine-tune it into eight instrument-specific experts (one per target stem), yielding a mixture-of-experts system where routing is determined by the known stem identity.


We train the restoration system in five stages, following the strategy proposed in~\cite{babaev2024finallyfastuniversalspeech} for stages 1-3:

\begin{itemize}
    \item \textbf{Stages 1–3:} Restore original music content (generalist warm-up), then introduce GAN training with feature matching, then introduce our music perceptual metrics.
    \item \textbf{Stage 4:} Focus on suppression of noise artifacts using additional augmentation, including gramophone noise.
    \item \textbf{Stage 5:} Fine-tune eight instrument-specific experts using instrument-filtered training pairs, produced by the separation model on separation training mixtures.    
\end{itemize}

The generalist restoration model is trained on the SonicMasterDataset~\cite{melechovsky2025sonicmaster}, after which we introduce a noise-focused training stage that augments the data with the Gramophone Record Noise Dataset~\cite{moliner2022two}. For expert specialization, we fine-tune on approximately $80$k excerpt-pair examples extracted from our 8-source separation data, where restoration inputs are generated by running separation inference to match test-time error characteristics. Restoration training was conducted on a single NVIDIA RTX PRO 6000 Blackwell GPU.

\section{Results and Limitations}

Table~\ref{tab:msr_perstem_obj_cpjku} summarizes our official MSR Challenge 2025 test-set results. The system achieves competitive objective performance (MMSNR 0.8329, Zimtohrli 0.0189, FAD 0.3814) and a system-level MOS of 3.5510. 
In addition to the official evaluation, we report here the average score on the MSRBench dataset~\cite{zang2025msrbench} for later iterations of our model, achieving $.638$ FAD and $2.338$ Multi-Mel-SNR (MMSNR).

A key limitation of our pipeline is sensitivity to noisy mixtures, especially live and historical recordings: in such cases the separator can fail to produce sufficiently faithful stem estimates, which in turn limits downstream restoration. We also observed that dataset mismatches and misalignments, particularly in large-scale collections such as RawStems, can bias restoration training and yield residual noise artifacts. Finally, time-varying effects embedded in training ground-truth stems (e.g., reverb, chorus, or delay) make a consistently “dry” target ambiguous, since the model is implicitly asked to remove effects that may be present in the reference. We therefore attribute most current failure modes to data quality and domain coverage, and plan to prioritize data curation, alignment verification, and effect-aware conditioning in future work.

\begin{table}[t]
\centering
\caption{Per-stem objective results for \textit{cp-jku} on the MSR Challenge 2025. The last row reports the unweighted mean.}
\label{tab:msr_perstem_obj_cpjku}
\setlength{\tabcolsep}{6pt}
\begin{tabular}{lccc}
\toprule
\textbf{Stem} & \textbf{MMSNR} & \textbf{Zimt} & \textbf{FAD} \\
\midrule
Bass                & 1.5486 & 0.0119 & 0.4490 \\
Drums               & 0.9552 & 0.0182 & 0.4912 \\
Guitars             & 1.2919 & 0.0194 & 0.5375 \\
Keyboards           & 0.7831 & 0.0174 & 0.7202 \\
Orchestral Elements & 0.5125 & 0.0208 & 0.7077 \\
Percussions         & 0.0994 & 0.0169 & 0.8479 \\
Synthesizers        & 0.6376 & 0.0271 & 0.9153 \\
Vocals              & 0.8351 & 0.0193 & 0.3092 \\
\midrule
\textbf{Average}    & \textbf{0.8329} & \textbf{0.0189} & \textbf{0.6223} \\
\bottomrule
\end{tabular}
\end{table}

\section{Conclusion and Future Work}

We presented a two-stage MSR system combining a curriculum-trained BS-RoFormer separator with instrument-specific HiFi++ GAN restoration experts. The separator is adapted using LoRA and extended from 4-stem to 8-stem prediction via staged training, while the restoration model is trained as a generalist and then specialized per instrument using separator-generated inputs to match test-time errors. 

\section*{Acknowledgments}

This work was supported by the European Research Council (ERC) under Horizon 2020 grant \#101019375 “Whither Music?”.

\bibliographystyle{IEEEbib}
\bibliography{strings,refs}

\end{document}